\documentclass[oldversion]{aa}
\usepackage{txfonts}
\usepackage{graphicx}

\begin{document}

   \title{Physical conditions in CaFe interstellar clouds}

   \subtitle{}

   \author{P. Gnaci\'nski,
           M. Krogulec
          }
   \authorrunning{P. Gnaci\'nski, M. Krogulec}
          
   \offprints{P. Gnaci\'nski}

   \institute{Institute of Theoretical Physics and Astrophysics,
              University of Gda\'nsk,
              ul. Wita Stwosza 57, 80-952 Gda\'nsk, Poland\\
              \email{pg@iftia.univ.gda.pl, fizmkr@univ.gda.pl}
         }

   \date{}

   \abstract{
     Interstellar clouds that exhibit strong Ca I and Fe I lines are so called CaFe clouds.
The ionisation equilibrium equations were used to model the column densities of Ca II, Ca I, K I, Na I, Fe I and Ti II in CaFe clouds. The chemical composition of CaFe clouds is that of the Solar System and no depletion of elements onto dust grains is seen. The CaFe clouds have high electron densities n$_e\approx1\ cm^{-3}$ that leads to high column densities of neutral Ca and Fe.
   }
   
   \maketitle

\section{Introduction}
  
\begin{figure}[bth]
   \centering
   \includegraphics[width=0.5\textwidth,viewport=1 1 686 470,clip]{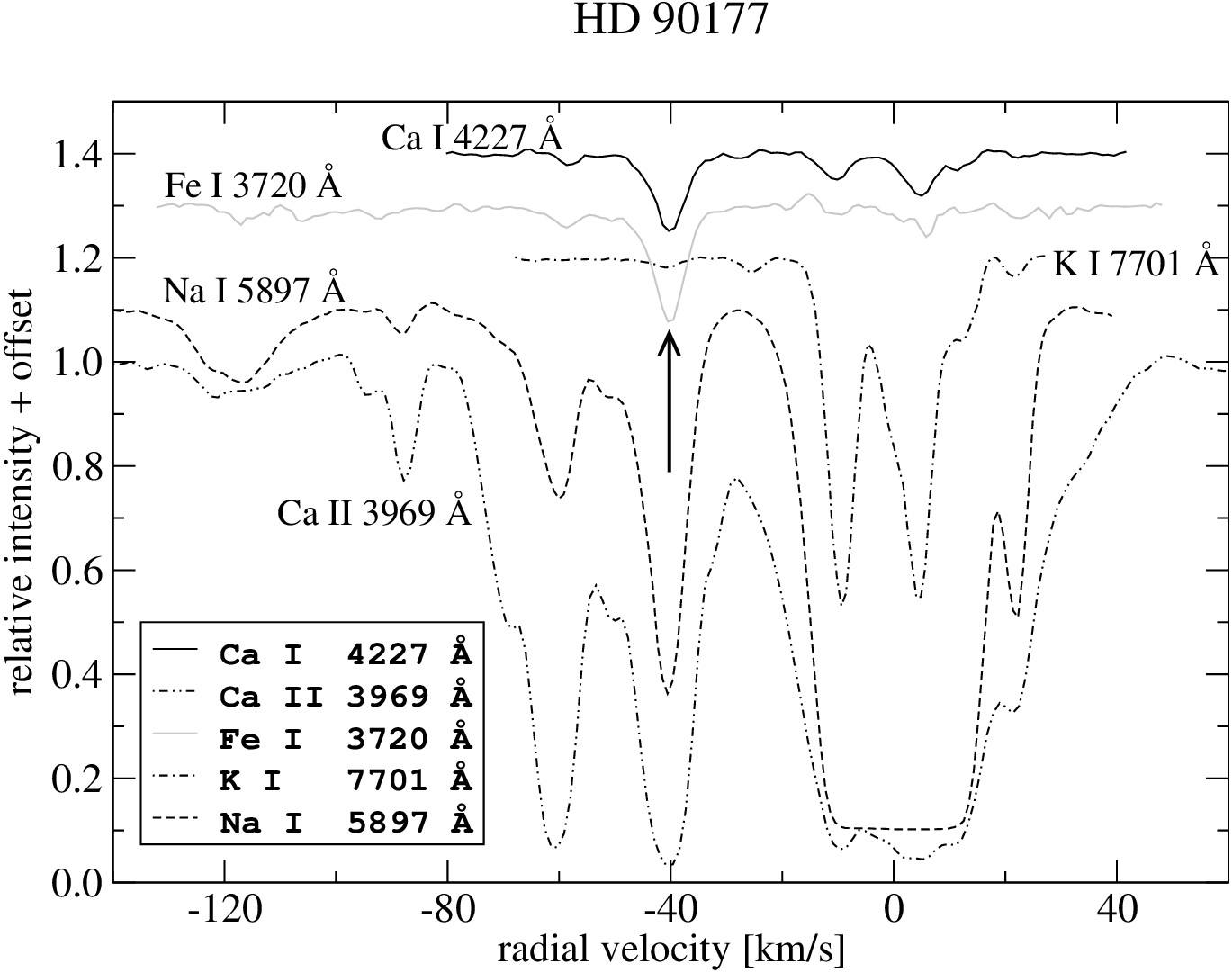}
      \caption{ Interstellar absorption lines in the direction of HD 90177. The arrow shows the position of the CaFe interstellar cloud.
              }
      \label{HD90177}
   \end{figure}

\begin{figure}[bth]
   \centering
   \includegraphics[width=0.5\textwidth,viewport=1 1 686 490,clip]{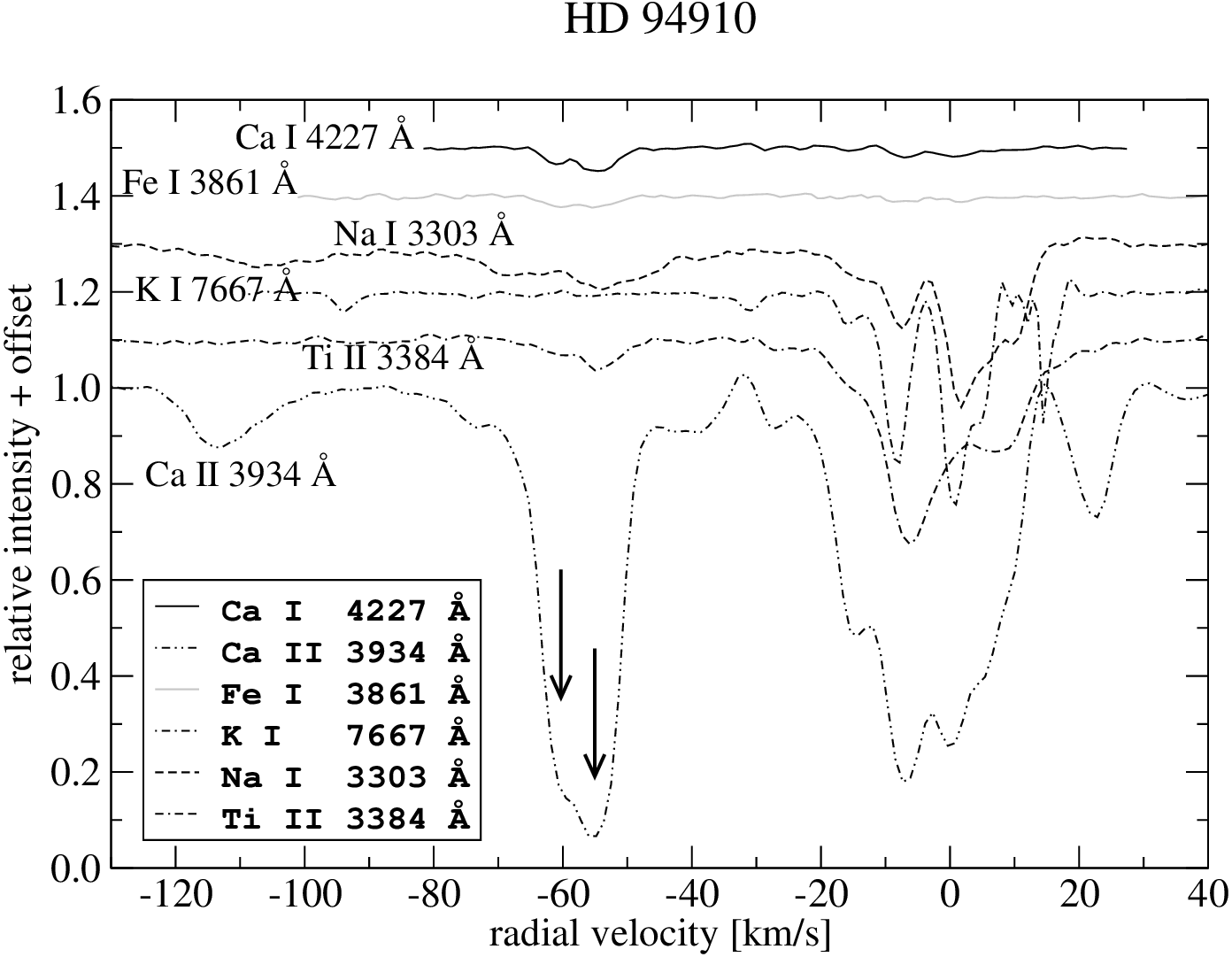}
      \caption{ Interstellar absorption lines in the direction of HD 94910. The arrows shows the positions of the CaFe interstellar clouds.
              }
      \label{HD94910}
   \end{figure}

Recently \cite{Bondar} have discovered a new type of interstellar clouds, so called CaFe clouds. These clouds are characterised by unusually strong lines of neutral Ca and Fe. This anomaly may be caused by (1) different chemical composition or (2) physical conditions in the CaFe clouds.
The two best examples of CaFe clouds, in the directions to HD 90177 and HD 94910, are shown on Fig. \ref{HD90177} and \ref{HD94910}.

  The Solar System composition is commonly used as a reference for chemical composition of matter. We have used the Solar System Abundances ($SSA(X)=\log X/H+12$) as summarised by \cite{Grevesse}. The deficit of elements in comparison to SSA is usually attributed to the depletion onto dust grains. 
  
\begin{table*}[bth]
\renewcommand{\thefootnote}{\alph{footnote}}
\caption{ Column densities (N in $cm^{-2}$) obtained from the UVES spectra. } 
\begin{tabular}{lccccccc}
\hline \hline
Star & v & N(CaII) 3970 \AA & N(CaI) 4228 \AA & N(KI) 7701 \AA & N(NaI) 5892 \AA & N(FeI) 3721 \AA & N(TiII) 3385 \AA \\
     & [$km/s$] & or 3935 \AA\ [$cm^{-2}$]  & [$cm^{-2}$]  & or 7667 \AA\ [$cm^{-2}$]  &  and 5898 \AA\ [$cm^{-2}$] & or 3861 \AA\ [$cm^{-2}$]   & or 3074 \AA\ [$cm^{-2}$]   \\
\hline 
HD 90177 & -40\footnotemark[3] &$ 6.2e12\pm7.8e11 $&$ 5.6e10\pm4.7e09 $&$ 2.2e10\pm8.5e09 $&$ 1.3e12\pm1.3e10 $&$ 3.0e12\pm1.6e11 $&$ 6.5e11\pm1.1e11 $\\
HD 94910 & -54\footnotemark[3] &$ 6.6e11\pm2.1e10 $&$ 1.9e10\pm1.6e09 $&$ <7e09 $&$ 2.6e11\pm4.2e10 $&$ 9.1e11\pm4.5e11 $&$ 1.2e11\pm4.9e10 $\\
HD 94910 &$ -7 $&$ 1.1e12\pm1.6e11 $&$ 5.0e09\pm8.7e08 $&$ 1.5e11\pm1.1e10 $&$ 3.5e12\pm8.1e10 $&$ 3.2e11\pm1.0e11 $&$ 1.2e12\pm1.3e11 $\\
HD 76341 &$ 20 $&$ 1.4e12\pm4.0e10 $&$ 5.6e09\pm1.6e09 $&$ 6.2e11\pm1.4e10 $&$ 7.2e13\pm1.0e12 $&$ 8.2e11\pm7.8e10 $&$ 4.9e11\pm2.8e10 $\\
HD 155806 &$ -3 $&$ 3.4e12\pm1.0e11 $&$ 1.1e10\pm4.4e09 $&$ 2.4e11\pm6.6e09 $&$ 2.4e13\pm1.7e12 $&$ 1.6e11\pm1.9e10 $&$ 6.3e11\pm4.6e10 $\\
HD 163800 &$ -6 $&$ 2.4e12\pm3.1e11 $&$ 2.5e10\pm1.9e09 $&$ 6.7e11\pm7.3e09 $&$ 6.2e13\pm7.5e11 $&$ 1.4e12\pm6.8e10 $&$ 8.3e11\pm5.0e10 $\\
\hline 
\end{tabular}
\newline \footnotemark[3]\footnotesize{CaFe cloud.} 
\label{gestKol}
\end{table*}


\begin{table*}
\renewcommand{\thefootnote}{\alph{footnote}}
\caption{Ionization equilibrium models.}
\begin{tabular}{lllllllll}
\hline \hline
star              & HD 90177 &  HD 94910 & HD 94910  & HD 76341 & HD 155806 & HD 163800 \\
v [km/s] & -40\footnotemark[3]& -54\footnotemark[3]& -7 & 20    & -3        & -6 \\
\hline
N(H) [$cm^{-2}$]  &$ 1.5^{+0.2}_{-0.2}e19 $&$ 2.1^{+0.4}_{-0.6}e18 $&$ 1.2^{+0.2}_{-0.1}e19 $&$ 7.9^{+0.3}_{-0.4}e18 $&$ 1.1^{+0.1}e19    $&$ 1.2^{+0.1}e19 $\\
n$_e$ [$cm^{-3}$] &$ 1.03^{+0.08}_{-0.05} $&$ 0.99^{+0.46}_{-0.21} $&$ 0.20^{+0.01}_{-0.01} $&$ 0.43^{+0.03}_{-0.02}   $&$ 0.60_{-0.12} $&$ 0.50_{-0.01} $\\
T$_e$ [$K$]       &$ 8400^{+300}_{-150}   $&$ 10500^{+600}_{-1200}  $&$ 8200^{+600}_{-700}   $&$ 8900^{+200}_{-200}   $&$ 5650_{-500}   $&$ 9000_{-300} $\\
\hline
N(H) from Ca [$cm^{-2}$] & 1.6e19 & 2.2e18 & 1.2e19  & 8.2e18  & 1.0e19  & 1.2e19 \\
N(H) from Ti [$cm^{-2}$] & 6.2e18 & 1.2e18 & 1.1e19  & 4.7e18  & ---     & 7.9e18 \\
\hline
\end{tabular}
\newline \footnotemark[3]\footnotesize{CaFe cloud.}
\label{model}
\end{table*}

\section{Column densities}

We have analysed interstellar clouds in the directions to HD 90177 and HD 94910 which are the best examples of CaFe clouds. In these two directions the CaFe clouds are well separated in the velocity scale from other interstellar clouds. 
For comparison we have used spectra of three stars with ordinary clouds: HD 76341, HD 155806 and HD 163800. The lines of sight to HD 94910, HD 76314, HD 155806 and HD 163800 were also analysed by \cite{Hunter}.

We have used publicly available spectra made with the UVES spectrograph at Paranal/Chile.
The spectra were made as part of 
"Library of High -- Resolution Spectra of Stars across the Hertzsprung -- Russell Diagram" \footnote{http://www.sc.eso.org/santiago/uvespop/} -- see \cite{Bagnulo} for details. The
spectral resolution R=$\lambda/\Delta\lambda$=80000 and the spectra are covering a broad range of wavelengths 3040 -- 10400 \AA. We have analysed interstellar absorption lines of neutral and ionised elements: Ca I, Ca II, Fe I, K I, Na I and Ti II.

The column densities of these elements were derived for individual clouds using the profile fitting technique. The absorption lines were fitted by Voigt profiles. The cloud velocities (v), Doppler broadening parameters (b) and column densities (N) for multiple absorption components were simultaneously fitted to the observed spectrum. The lines of Na doublet (at 5892 \AA\ and 5898 \AA) were also fitted simultaneously - v, b and N were common for both lines in the doublet. The wavelengths, oscillator strengths (f) and natural damping constants ($\Gamma$) were adopted from \cite{Morton}.
The derived column densities are presented in Table \ref{gestKol}.

\section{Ionisation equilibrium}

We have used the ionisation equilibrium equation for the Ca, Fe, Na, K and Ti elements to determine the physical conditions (electron density and temperature) in CaFe clouds. 
  In the case of Ca the equation of ionisation equilibrium is the following:
  \begin{equation}
     \frac{n_{e}N(Ca\: II)}{N(Ca\: I)}=\frac{\Gamma(Ca_{12})+n_{e}C(Ca_{12})}{\alpha_{rad}(Ca_{21})+\alpha_{die}(Ca_{21})}
  \label{rownowaga}
  \end{equation}
     where $N(Ca\:II)$ and $N(Ca\:I)\ [cm^{-2}]$ are the column densities of ionised and neutral Ca, $\alpha_{rad}(Ca_{21})\ [cm^3/s]$ is the radiative recombination coefficient, $\alpha_{die}(Ca_{21})\ [cm^3/s]$ is the dielectronic recombination rate, $\Gamma(Ca_{12})\ [1/s]$ is the ionisation rate of Ca I by UV photons, and $C(Ca_{12})\ [cm^3/s]$ is the collisional ionisation rate. Similar equations were used for other elements.
 The $\alpha_{rad}$, $\alpha_{die}$ and $C$ parameters depends on the electron temperature T$_e$.

  The $\Gamma$ coefficients were adopted from the WJ2 model (\cite{Boer}). The recombination coefficients ($\alpha_{rad}$ and $\alpha_{die}$) and the collisional ionisation rate coefficient ($C$) for the Ca and Fe elements 
were adopted from \cite{Shull}. For K and Na these parameters were adopted from  \cite{Landini1991} and \cite{Landini1990}. 
For Ti $\alpha_{rad}$ was adopted from \cite{Badnell} and $\alpha_{die}$ comes from \cite{Mazzotta}. The collisional ionisation rate coefficient ($C$) for Ti was adopted from \cite{Voronov}.

  The CH and CN molecules are not present in CaFe clouds (\cite{Bondar}). The CH column 
density is directly proportional to the column density of the H$_2$ molecule (eg.
\cite{Federman}, \cite{Danks}, \cite{GnaKroKre}). Furthermore, the H$_2$ molecule is known to be formed on dust grains.
We infer that the absence of simple molecules is caused by grains completely missing in CaFe clouds. \cite{Weingartner} have proposed recombinations on polycyclic aromatic hydrocarbons (PAHs) and on dust grains to play an important role in ion recombinations in cold neutral medium. Since we do not observe even the simplest molecules in CaFe clouds we
do not consider the recombinations on PAHs or dust grains here.

 The second equation necessary to determine $N(Ca\:I)$ and $N(Ca\:II)$ in our model was the conservation of the number of Ca atoms:
  \begin{equation}
     N(Ca\:I)+N(Ca\:II)+N(Ca\:III)=N(H)\cdot 10^{SSA(Ca)-12}
  \end{equation}
  where $N(Ca\:III)$ was calculated from the ionisation equilibrium with observed $N(Ca\:II)$. We have used the Solar System Abundances (SSA) as given by \cite{Grevesse}.
  Other elements (eg. Fe) have the second ionisation potential larger or close to 13.6 eV, and are present in the interstellar medium only in two ionisation stages:
  \begin{equation}
     N(Fe\:I)+N(Fe\:II)=N(H)\cdot 10^{SSA(Fe)-12}
  \end{equation}
  
    The hydrogen column density N(H), the electron density n$_e$ and the electron temperature T$_e$ were treated as free parameters. 
We have used the {\it `amoeba'} procedure \cite{Press} to find the minimum of the $\chi$
function:
  \begin{equation}
     \chi^2=\sum_{X} \left(\frac{N(X)_{model}}{N(X)_{obs}}-1\right)^2
  \end{equation}
  where $X$=Ca I, Ca II, Na I, Fe I, K I and Ti II.
 
\begin{figure}[bth]
   \centering
   \includegraphics[width=0.5\textwidth,viewport=1 1 537 770,clip]{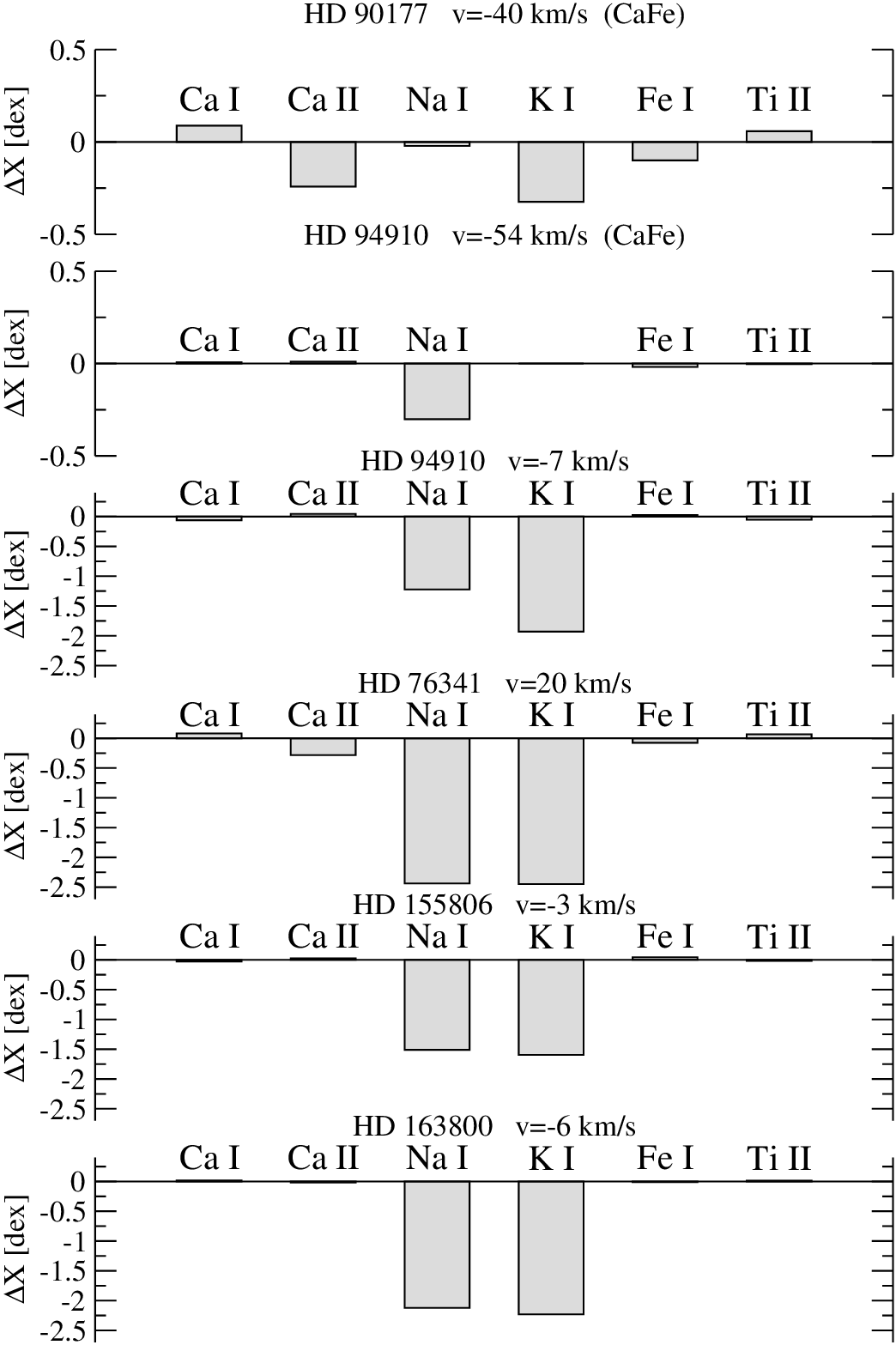}
      \caption{ The difference $\Delta$X=$\log N(X)_{model} - \log N(X)_{obs}$ between
      calculated column densities and the observed ones. Note different y-axis scale for
      CaFe and non-CaFe clouds.
              }
      \label{Errors}
   \end{figure}

 The model that fits the observed column densities is presented in Table \ref{model}.
N(H) is the column density of hydrogen in a single analysed cloud. Physical condition are described by
the electron density n$_e$ and the electron temperature T$_e$. For comparison we have also shown the hydrogen column densities obtained from Ti II (assuming $N(Ti\:II)=N(Ti)$) and from Ca II ( $N(Ca\:I)_{obs}+N(Ca\:II)_{obs}+N(Ca\:III)_{model}=N(Ca)$) assuming Solar System abundances. The differences between the column densities obtained in our model and the observed ones $\Delta$X=$\log N(X)_{model} - \log N(X)_{obs}$ are shown on Fig. \ref{Errors}.

  The errors of N(H), n$_e$ and T$_e$ were calculated by modifying the column densities by the error presented in Table \ref{gestKol}. First we have added the errors to the column densities of ions and subtracted the error from the column densities of neutral elements. In a second run we have added the error to column densities of neutral species and subtracted from ions column densities. In a case of interstellar clouds in the direction to HD 155806 and HD 163800 we could not achieve convergence in the second case.

\section{Discussion}

The absence of molecules in CaFe clouds imply that no dust is present in this clouds. This conclusion is confirmed by comparison of element abundances in CaFe clouds and in our model.
  Elements with low condensation temperature like Na or K (T$_C\approx1000\ K$) do not show any depletion in the CaFe clouds. In fact the observed column densities of Na I and K I are even larger than column densities derived from the best--fit model (Fig. \ref{Errors}).
  The depletion of atoms due to condensation onto dust grains is typically -0.2 to -1.5 dex (\cite{GnaKro}), while the highest deficit of observed abundance in our data is 0.088 dex for Ca I in the CaFe cloud to HD 90177.
  Therefore we conclude that the CaFe clouds have composition similar to the Solar System abundances and no depletion onto dust grains is seen in the CaFe clouds.
  
  Both CaFe clouds analysed here have large electron density n$_e\approx1\ cm^{-3}$.
  In clouds analysed by \cite{Gna} the electron density is in the range 0.01 -- 2.5 $cm^{-3}$. The electron densities in the CaFe clouds are higher than electron densities to all 11 sight lines presented by \cite{Black}.
  
  It is interesting to note that our model fails to fit Na I and K I in ordinary (non CaFe)
clouds. The observed column densities of Na I and K I are always higher by 1--3 orders of magnitude than the theoretical ones. It is possible that a single set (n$_e$, T$_e$) can not describe the conditions in a non--CaFe cloud, because in the cloud's core the ionising photons are shielded by dust, and the conditions in the cloud's core differ substantially from that on the cloud surface. The CaFe clouds seems to be very homogenous in comparison to ordinary clouds, since we can fit the observed column densities with a single set of n$_e$ and T$_e$. In CaFe clouds we do not see the CH and CH$+$ molecules (\cite{Bondar}). The fact that we do not observe molecules in CaFe clouds support the thesis that there are no cold cores in such clouds.

\section{Conclusions}

 The results can be recapitulated as follows:
   \begin{enumerate}
      \item The CaFe interstellar clouds have composition similar to the Solar System elements abundance.
      \item There is no depletion of elements onto dust grains in CaFe clouds.
      \item The electron density n$_e$ is high in the CaFe clouds (n$_e\approx1\ cm^{-3}$).    
      \item The CaFe clouds seems to be thin clouds totally penetrated by the UV radiation.
   \end{enumerate}

\begin{acknowledgements}
    This research was based on spectra obtained by the UVES Paranal Observatory Project (ESO DDT Program ID 266.D-5655, see \cite{Bagnulo}). We acknowledge the financial support from the University of Gda\'nsk (grant BW/5400-5-0305-7).
    
\end{acknowledgements}

\end{document}